# Novel sparse PCA method via Runge Kutta numerical method(s) for face recognition

Loc Hoang Tran, Luong Anh Tuan Nguyen

Faculty of Information Technology - Vietnam Aviation Academy

locth@vaa.edu.vn, nlatuan@vaa.edu.vn

**Abstract:** Face recognition is a crucial topic in data science and biometric security, with applications spanning military, finance, and retail industries. This paper explores the implementation of sparse Principal Component Analysis (PCA) using the Proximal Gradient method (also known as ISTA) and the Runge–Kutta numerical methods. To address the face recognition problem, we integrate sparse PCA with either the k-nearest neighbor method or the kernel ridge regression method. Experimental results demonstrate that combining sparse PCA—solved via the Proximal Gradient method or the Runge–Kutta numerical approach—with a classification system yields higher accuracy compared to standard PCA. Additionally, we observe that the Runge–Kutta-based sparse PCA computation consistently outperforms the Proximal Gradient method in terms of speed.

**1. Introduction**

Face recognition is a form of biometric security with wide-ranging applications in fields such as finance, military, and retail. In the 1990s, computer scientists introduced the Eigenface approach [1], which utilized Principal Component Analysis (PCA) [2, 3] for dimensionality reduction to filter out noise and redundant information, thereby speeding up face recognition systems. PCA has also been applied in various pattern recognition domains, including speech recognition [4, 5]. The Eigenface method typically paired PCA with the k-nearest neighbor algorithm for classifying faces [6]. Later, machine learning techniques [7, 8, 9, 10, 11, 12] like Support Vector Machines (SVM) became state-of-the-art for such tasks between 2000 and 2010, with kernel ridge regression often viewed as a simpler form of SVM. Although deep convolutional neural networks now dominate image recognition, our study does not employ them due to constraints in time and space.

Notably, PCA suffers from two key limitations: its principal components are dense, meaning each component is a linear combination of all features, and there is no inherent sparsity in the loading vectors. To address these issues, various sparse PCA methods have been proposed [13, 14, 15, 16]. In our paper, we introduce two novel approaches for sparse PCA that leverage the Proximal Gradient (ISTA) method [17, 18, 19] and Runge–Kutta numerical techniques [21]. Finally, we integrate these sparse PCA methods with both the k-nearest neighbor and kernel ridge regression classifiers to tackle the face recognition problem.

The paper is structured as follows: Section 2 details the sparse PCA algorithm using the proximal gradient method, Section 3 explains the sparse PCA algorithm based on Runge–Kutta methods, Section 4 presents experimental applications combining these sparse PCA approaches with k-nearest neighbor and kernel ridge regression on a face dataset, and Section 5 concludes with future research directions in face recognition.

## 2. The proximal gradient method

Suppose that we would like to solve the unconstrained optimization problem like the following:

$$min_z f(z)$$

One of the simplest methods solving the above unconstrained optimization problem is the gradient method. This gradient method solves the above optimization as follows:

$$z_0 \in R^n, z_{k+1} = z_k - t_{k+1} \nabla f(z_k), \text{ where } t_{k+1} \text{ is the suitable step size.}$$

The idea of this gradient method can also be applied to the $l_1$-regularization problem like the following:

Suppose that we are given a function $f$ that can be decomposed into two functions as follows:

$$f(x) = g(x) + \lambda h(x)$$

For this decomposable function $f$, we want to find the solution of the following optimization problem

$$min_z f(z) = g(z) + \lambda h(z)$$

Please note that function $g$ is differentiable.

So, the solution of this optimization problem is:

$$x^+ = argmin_z g(x) + \nabla g(x)^T (z - x) + \frac{1}{2t}||z - x||_2^2 + \lambda h(z)$$

$$= argmin_z \frac{1}{2t}||z - (x - t\nabla g(x))||_2^2 + \lambda h(z)$$

$$= prox_{h,\lambda t}(x - t\nabla g(x))$$

Please note that $\nabla^2 g(x)$ is replaced by $\frac{1}{t}I$.

This method can also be called the ISTA method. In the next section, we will present the Runge Kutta numerical method(s) used to solve this specific sparse PCA problem and are a lot faster than the ISTA method.

In our sparse PCA problem, $g(x) = -x^T D^T Dx$ and $h(x) = ||x||_1$.

So, we know that: $\nabla g(x) = -2D^T Dx$.

Hence the solution of sparse PCA problem is:

$$x^+ = prox_{h,\lambda t}(x + 2tD^T Dx)$$

$$= prox_{h,\lambda t}((I + 2tD^T D)x)$$

Finally, in detail, the proximal operator is defined like the following:

$$prox_{h,\lambda t}(x - t\nabla g(x))_i = (|x + 2tD^T Dx|_i - \lambda t)_+ sign((x + 2tD^T Dx)_i)$$

We repeat the above formula until it converges to final solution.

In the next part, we will give the algorithm of one pass of the proximal gradient method (i.e. the

ISTA method).

Algorithm 1: One pass of proximal gradient method

1. Start with $x_k$.
2. Compute $x_{k+1} = prox_{h,\lambda t}((I + 2tD^TD)x_k)$.
3. $x_{k+1}$ will be served as the input for the next pass of the ISTA method.

## 3. The Runge Kutta numerical method(s)

First, we see that there is another view from which we can derive the gradient descent procedure. Say we have a differential equation of the form:

$$\frac{dx}{dt} = u(x)$$

We can use Euler's method to solve this equation by choosing a starting point $x_0$ and iteratively applying the following approximation for a step size $t$ until we reach a steady state $x_s$ such that $u(x_s) = 0$:

$$x_{k+1} = x_k + t\frac{dx}{dt} = x_k + tu(x_k)$$

Now let $u(x) = -\nabla g(x)$. Applying the Euler's method to the above equation, we acquire:

$$x_{k+1} = x_k - t\nabla g(x_k),$$

which is exactly the update step for gradient descent! Note that this differential equation reaches a steady state $x_s$ such that $\nabla g(x_s) = 0$. This must be an extremum of $g$.

Inspired by this idea (using the Euler's method to acquire $x_{k+1}$), we can employ the Runge Kutta numerical method(s) to obtain $x_{k+1}$ like the following:

a. Coarse Runge Kutta method:

$$x_{k+\frac{1}{2}} = x_k - t\nabla g(x_k)$$

$$x_{k+1} = x_k - t\nabla g\left(x_{k+\frac{1}{2}}\right)$$

Hence, we easily see that $x_{k+1} = x_k - t\nabla g(x_k - t\nabla g(x_k))$.

b. Fourth order Runge Kutta method:

$$x_{k+1} = x_k + \frac{t}{6}(s_1 + 2s_2 + 2s_3 + s_4),$$

where:

$$s_1 = -\nabla g(x_k)$$
$$s_2 = -\nabla g\left(x_k + t\frac{s_1}{2}\right)$$
$$s_3 = -\nabla g\left(x_k + t\frac{s_2}{2}\right)$$
$$s_4 = -\nabla g(x_k + ts_3)$$

After acquiring $x_{k+1}$, we can compute $prox_{h,\lambda t}(x_{k+1})$ to get the novel version of sparse PCA method (by using the Runge Kutta numerical method(s)).

Finally, please note that $g(x) = -x^TD^TDx$.

## 4. Experiments

In this study, we use a training set composed of 120 face images collected from 15 individuals (with 8 images per person) and a testing set of 45 face images from the same group, as obtained from [20]. Each face image is initially represented as a matrix; we then convert each matrix into a single row vector in $R^{1*1024}$ by concatenating its rows. These row vectors serve as feature inputs for both the k-nearest neighbor method and the kernel ridge regression method. Subsequently, we apply both PCA and sparse PCA to the training and testing sets to reduce the dimensionality of the face data, and then we perform classification using the k-nearest neighbor and kernel ridge regression methods on the transformed features. In our experiments, we evaluate the accuracy of the k-nearest neighbor and kernel ridge regression methods.

All experiments were executed in Python on Google Colab using an NVIDIA Tesla K80 GPU with 12 GB of RAM. The performance outcomes, including the accuracies of our proposed methods, are summarized in Tables 1 and 2, while Table 3 presents the time complexities for computing the sparse PCA via the proximal gradient and Runge–Kutta numerical methods.

Table 1: **Accuracies** of the nearest-neighbor method, the combination of PCA method and the nearest-neighbor method, and the combination of sparse PCA method and the nearest-neighbormethod

| | Accuracy |
|---|---|
| PCA (d=20) + $k$ nearest neighbor method | 0.64 |
| PCA (d=30) + $k$ nearest neighbor method | 0.67 |
| PCA (d=40) + $k$ nearest neighbor method | 0.64 |
| PCA (d=50) + $k$ nearest neighbor method | 0.67 |
| PCA (d=60) + $k$ nearest neighbor method | 0.67 |
| ISTA sparse PCA (d=20) + $k$ nearest neighbor method | 0.69 |
| ISTA sparse PCA (d=30) + $k$ nearest neighbor method | 0.69 |
| ISTA sparse PCA (d=40) + $k$ nearest neighbor method | 0.69 |
| ISTA sparse PCA (d=50) + $k$ nearest neighbor method | 0.69 |
| ISTA sparse PCA (d=60) + $k$ nearest neighbor method | 0.71 |
| Coarse Runge Kutta sparse PCA (d=20) + $k$ nearest neighbor method | 0.71 |
| Coarse Runge Kutta sparse PCA (d=30) + $k$ nearest neighbor method | **0.73** |
| Coarse Runge Kutta sparse PCA (d=40) + $k$ nearest neighbor method | 0.71 |
| Coarse Runge Kutta sparse PCA (d=50) + $k$ nearest neighbor method | 0.71 |
| Coarse Runge Kutta sparse PCA (d=60) + $k$ nearest neighbor method | 0.71 |
| Fourth order Runge Kutta sparse PCA (d=20) + $k$ nearest neighbor method | 0.71 |
| Fourth order Runge Kutta sparse PCA (d=30) + $k$ nearest neighbor method | 0.71 |
| Fourth order Runge Kutta sparse PCA (d=40) + $k$ nearest neighbor method | 0.71 |
| Fourth order Runge Kutta sparse PCA (d=50) + $k$ nearest neighbor method | 0.71 |
| Fourth order Runge Kutta sparse PCA (d=60) + $k$ nearest neighbor method | 0.69 |

Table 2: **Accuracies** of the kernel ridge regression method, the combination of PCA method andthe

kernel ridge regression method, and the combination of sparse PCA method and the kernel ridge regression method

| Accuracy | |
|---|---|
| PCA (d=20) + kernel ridge regression method | 0.80 |
| PCA (d=30) + kernel ridge regression method | 0.84 |
| PCA (d=40) + kernel ridge regression method | 0.84 |
| PCA (d=50) + kernel ridge regression method | 0.80 |
| PCA (d=60) + kernel ridge regression method | 0.84 |
| ISTA sparse PCA (d=20) + kernel ridge regression method | 0.80 |
| ISTA sparse PCA (d=30) + kernel ridge regression method | 0.84 |
| ISTA sparse PCA (d=40) + kernel ridge regression method | 0.84 |
| ISTA sparse PCA (d=50) + kernel ridge regression method | 0.84 |
| ISTA sparse PCA (d=60) + kernel ridge regression method | 0.84 |
| Coarse Runge Kutta sparse PCA (d=20) + kernel ridge regression method | 0.84 |
| Coarse Runge Kutta sparse PCA (d=30) + kernel ridge regression method | 0.84 |
| Coarse Runge Kutta sparse PCA (d=40) + kernel ridge regression method | **0.87** |
| Coarse Runge Kutta sparse PCA (d=50) + kernel ridge regression method | **0.87** |
| Coarse Runge Kutta sparse PCA (d=60) + kernel ridge regression method | **0.87** |
| Fourth order Runge Kutta sparse PCA (d=20) + kernel ridge regression method | **0.87** |
| Fourth order Runge Kutta sparse PCA (d=30) + kernel ridge regression method | **0.87** |
| Fourth order Runge Kutta sparse PCA (d=40) + kernel ridge regression method | 0.84 |
| Fourth order Runge Kutta sparse PCA (d=50) + kernel ridge regression method | **0.87** |
| Fourth order Runge Kutta sparse PCA (d=60) + kernel ridge regression method | **0.87** |

Table 3: **The time complexities** of the processes computing the sparse PCA using the proximal gradient method (ISTA method) and the Runge Kutta numerical methods

| Time complexities (seconds) | |
|---|---|
| ISTA sparse PCA (d=20) | 4.18 |
| ISTA sparse PCA (d=30) | 3.48 |
| ISTA sparse PCA (d=40) | 5.35 |
| ISTA sparse PCA (d=50) | 3.24 |
| ISTA sparse PCA (d=60) | 4.53 |
| Coarse Runge Kutta sparse PCA (d=20) | 1.96 |
| Coarse Runge Kutta sparse PCA (d=30) | 1.99 |
| Coarse Runge Kutta sparse PCA (d=40) | 2.01 |
| Coarse Runge Kutta sparse PCA (d=50) | 1.98 |
| Coarse Runge Kutta sparse PCA (d=60) | 1.95 |

| Fourth order Runge Kutta sparse PCA (d=20) | 1.83 |
| Fourth order Runge Kutta sparse PCA (d=30) | 1.81 |
| Fourth order Runge Kutta sparse PCA (d=40) | 1.87 |
| Fourth order Runge Kutta sparse PCA (d=50) | 1.80 |
| Fourth order Runge Kutta sparse PCA (d=60) | 1.85 |

From the above tables 1 and 2, we recognize the accuracies of the combination of the sparse PCA method (implemented by the proximal gradient method and the Runge Kutta numerical method(s)) and one specific classification system are higher than the accuracy of the combination of the PCA method and one specific classification system.

Finally, from the experiments, the process computing the sparse PCA algorithm using Runge Kutta numerical method(s) is faster than the process computing the sparse PCA algorithm using proximal gradient method.

4. **Conclusions**

In this paper, we introduce a new sparse PCA algorithm that is solved using both the Proximal Gradient method (also known as ISTA) and various Runge–Kutta numerical methods. Our experiments show that when sparse PCA is combined with the k-nearest neighbor or kernel ridge regression classifiers, the resulting accuracy is higher than that achieved by pairing standard PCA with these classifiers. Additionally, our findings indicate that the fourth-order Runge–Kutta implementation of sparse PCA runs faster than the Proximal Gradient approach.

Looking ahead, we plan to explore sparse PCA formulations solved by other numerical methods typically used for ordinary differential equations, and we aim to integrate sparse PCA with advanced deep learning architectures, such as VGG-based convolutional neural networks, to address recognition tasks not only in images but also in speech, text, and beyond.